\documentclass[twocolumn]{aastex631}
\received{\today}
\usepackage{tabularx}
\graphicspath{{figures/}}
\usepackage{aas_macros}
\usepackage{lipsum}
\usepackage{physics}
\usepackage{multirow}
\usepackage{xspace}
\usepackage{natbib}
\usepackage{fontawesome5}
\usepackage{xcolor}
\usepackage{amsmath}
\usepackage{wrapfig}
\usepackage{physics}
\usepackage{booktabs} 
\usepackage{bm}
\usepackage[figuresright]{rotating} 
\usepackage[version=4]{mhchem}
\setcitestyle{numbers,square,comma,sort&compress}
\usepackage[hang,flushmargin]{footmisc}

\newcommand{\bfX}{{\bf X}}

\newcommand{\bfk}{{\bf k}}

\makeatletter

\newcommand{\checknextarg}{\@ifnextchar\bgroup{\gobblenextarg}{}}
\newcommand{\gobblenextarg}[1]{\,\mathrm{#1}\@ifnextchar\bgroup{\gobblenextarg}{}}
\makeatother

\begin{document}

\title{\Large Two bits about lossy compression: On the limits of compression in cosmology}

% affiliations
\newcommand{\UW}{Department of Astronomy, University of Washington, Seattle, WA, 98195}

\author{Hurum Maksora Tohfa}
\affiliation{\UW}
\author{Matthew McQuinn}
\affiliation{\UW}

\begin{abstract}

%Astronomy is in an era where large numbers of simulations are being run for parameter inference. It can be challenging to store these datasets, and downloading them from public repositories can be extremely slow. While the outputs of cosmological simulations are typically stored as arrays of 32-bit floating point numbers, there is little reason to save data to better than percent level accuracy on the pixel scale, both because of numerical noise and uncertainties in the astrophysical inputs. The statistical translational invariance of cosmological fields and the approximate Gaussian distribution of mode amplitudes, even on nonlinear scales, enable a clean application of seminal results from compression theory. 

Astronomy is in an era of enormous sky surveys and of the large simulation suites needed to interpret them, both costly to store and slow to share. Simulation outputs are stored as 32-bit floats, yet numerical noise and astrophysical uncertainties make better than percent-level pixel accuracy unnecessary. Because cosmological fields are statistically homogeneous with nearly Gaussian mode amplitudes, classic rate-distortion results apply directly, and non-Gaussian structure permits further compression. We use the scientific compression package SZ3, and a neural compressor that fixes the quantization and learns the probability of each quantization bin with an autoregressive transformer. SZ3 beats the Gaussian approach only where the field is smooth on the pixel scale or its values pile up at a single point, while the neural approach matches or beats SZ3 on every field we consider and is 9 to 27\% below the Gaussian-optimal coder at fixed distortion, with the largest margin where the field is most non-Gaussian. Because the quantizer, not the network, sets the error, a poorly trained model can waste bits but never cost accuracy. A model trained only on weak lensing convergence maps transfers to $N$-body density fields without retraining, suggesting it has learned generic properties of cosmological structure rather than features of one dataset. Eulerian grids need only 1-4 bits per pixel (bpp) at percent-level accuracy, and particle displacements and velocities 5-6 at the precisions they require. The approach carries over to observational data: on Rubin Observatory Data Preview 1 coadds, with the quantization step set to a quarter of the background noise, the fine-tuned transformer needs 4.0 bpp, 27\% fewer than the Gaussian coder.

\end{abstract}

\section{Introduction}
\label{sec:intro}

Modern cosmological inference requires large suites of simulations, each with resolution elements numbering in the billions \citep[e.g.][]{vogelsbergersimreview}. The suite of simulations required for analysis must also span all potential values for cosmological parameters and often other astrophysical parameters \citep{2019MNRAS.488.4440A, 2020PNAS..11730055C, 2023arXiv230910270M}, resulting in datasets that range from terabytes to petabytes and, hence, are difficult to store, analyze, and transfer. Yet, their outputs are stored as 32-bit floating-point numbers, far more precision than would be needed
for any conceivable analysis and well beyond the accuracy of the simulations themselves.

%We show that percent-level accuracy at the resolution scale of the data set, already likely smaller than the baryonic uncertainties and simulation noise at these scale, requires only a few bits per floating-point value. A Fourier-space scheme that treats modes as independent and Gaussian achieves this at $2$--$4$ bpp, and the penalty incurred because cosmological fields are not independent Gaussian random fields is under $0.1$~bit/pixel. Because a Gaussian field is the highest-entropy field at fixed power spectrum, and hence the least compressible, the remaining gains must come from non-Gaussian correlations between nearly co-spatial quantities. We explore off-the-shelf SZ compressor \citep{SZcompressor}, which interpolates between points, gives at best modest gains beyond the Gaussian model. We then consider neural compressors that learn the probability distribution of the data, ranging from a network that learns only the per-mode bit allocation of the Fourier scheme of \S\ref{sec:gaussian}, to end-to-end learned transform coding \citep{2016arXiv161101704B, 2022arXiv220206533Y}, to an autoregressive transformer entropy model \citep{2017arXiv170603762V} applied at fixed quantization.

This Letter is one of only a few studies of lossy compression in cosmology, following \citep{Pulido_2019}, \citep{2020arXiv200400224J, 2021arXiv210400178J} and \citep{guppy}. Previous work used systematic errors on the power spectrum as the metric for how much loss could be tolerated \citep{Pulido_2019, 2020arXiv200400224J, 2021arXiv210400178J}; our perspective is closer to \citep{guppy}, that it would be a waste to spend millions of processing unit hours on a simulation without storing pixel-level outputs to sufficient precision for any conceivable application. We also check that the power spectra of these compressed fields are well below the 1\% error required for precision cosmology.  

Rather than applying established compression algorithms as in most of these previous works, we develop the theory for achievable compressions on the cosmological data sets discussed in \S\ref{sec:numericalsims} and explore algorithmic trade-offs in what follows.  This includes a compression algorithm that is optimal for a Gaussian random field (\S\ref{sec:gaussian}), a state-of-the-art scientific software that compresses with interpolation (\S\ref{sec:sz3}), and a deep learning based method (\S\ref{sec:neural}).   As different schemes use different informational properties of the data set to compress, this study also has bearing on where cosmological information lies.

%This Letter is organized as follows. We describe the datasets in \S\ref{sec:numericalsims}. In \S\ref{sec:gaussian} we review the Gaussian compression scheme, in \S\ref{sec:sz3} the state of the art scientific compressor SZ3, and in \S\ref{sec:neural} we introduce our deep learning based compressor and benchmark all three. We conclude in \S\ref{sec:conclusion}..

\section{numerical datasets}
\label{sec:numericalsims}

We consider three cosmological simulation datasets to understand what compressions may be possible, chosen to span the different types of cosmological data.  Namely,
\begin{description}

\item [2D weak lensing convergence maps from  \citep{2019MNRAS.490.1843R}]  This dataset set consists of 92 different cosmologies (scanning in $\sigma_8$ and $\Omega_m$) and in each cosmology 512 different random seeds, with each map being a $1024^2$ image of a $3.5\times 3.5$ degree$^2$ patch of the sky. While this dataset is only a terabyte, downloading it took us a week.
%; compression could facilitate use, and including it in this study affords an algorithmic warm up to the 3D datasets.

\item[Eulerian-gridded Bolshoi simulation from \citep{2011ApJ...740..102K}\footnote{ \url{https://www.cosmosim.org}}]  The $z=0$ density field in $512^3$ resolutions, from this $250/h$\;Mpc N-body simulation. These 3D Eulerian grids are calculated from the $2048^3$ particles using cloud-in-cell interpolation (CIC). 

\item[CAMELS $N$-body simulation \citep{camels, camels_dr}]  A dark matter-only run with $256^3$ particles in a $(25\,h^{-1}\,{\rm Mpc})^3$ periodic box, evolved from $z=127$ to $z=0$. We take the particle displacements and velocities at $z=0$, placed on the $256^3$ Lagrangian grid recovered from the particle identifiers \citep{camels,camels_dr}.\footnote{This does require a sort of simulation outputs by particle ID, which may be challenging for the largest simulations.  It would be interesting to investigate the success of compression on the semi-sorted outputs of most $N$-body codes.}
 
\end{description}
Although these are relatively small simulations or maps, the size is not important for understanding how effectively cosmological simulations can be compressed. 

While we expect a $1\%$ precision at the pixel scale to be sufficient for most simulation outputs, cosmological displacements are likely the exception, as one wants to capture the smallest structures the simulation resolves, which are set by the gravitational softening length. \citet{guppy} reconstruct each particle at a random point within its cell, so the RMS error is the cell width over $\sqrt{6}$. For the CAMELS force softening, $\epsilon = 2.4\,h^{-1}$\,kpc or $1/40$ of the mean interparticle spacing,\footnote{\url{https://camels.readthedocs.io}, page \textit{Codes}.} and the $2.5\,h^{-1}$\, Mpc standard deviation of the $x$ displacement, a distortion of $10^{-3}$ corresponds to a cell of about $2.5\epsilon$. This is coarser than the $2\epsilon$ at which they find percent-level biases in low-mass halo properties. For work on halo scales we also report $10^{-4}$, a cell of $0.25\epsilon$, finer than the $\epsilon$ they recommend.

%They find that a cell of $2\epsilon$ biases halo properties at only the percent level, so $10^{-3}$ is adequate for studies of the larger-scale distribution of matter. For work on halo scales we also report $10^{-4}$, a cell of $0.25\epsilon$, finer than the $\epsilon$ they recommend.

%As you want to locate a halo to better than a percent, the particle displacement needs to be stored with more precision to resolve cosmological structures. 
%For a typical cosmological displacement ($\approx 10$~Mpc at low redshift).\footnote{This $\approx 10$~Mpc would only apply if this displacement is captured, which requires box sizes $\gg 300$Mpc}, this works out to $10^{-3}$ precision for our CAMELS simulation.

\section{Compression algorithms}

We start with the classic result from \citep{shannon} that the minimum average code length for a quantized value is its entropy. For independent pixels, the entropy is
\begin{equation}
{\rm bpp} = -\sum_i p_i \log_2 p_i,
\label{eqn:shannon}
\end{equation}
where $p_i$ is the probability of falling in the $i^{\rm th}$ quantization bin. More generally, for a vector $\bfX$ of floating-point numbers with $P_{Q_i}(\bfX)$ the probability that $\bfX$ falls in bin $Q_i$, the entropy is $-\sum_i P_{Q_i}(\bfX)\log_2 P_{Q_i}(\bfX)$; equation~(\ref{eqn:shannon}) is the one-element case. Thus, compression boils down to how well one can model the probability distribution $P_{Q_i}(\bfX)$. The three methods below differ in how they model $P_{Q_i}(\bfX)$: as independent Gaussian Fourier modes (\S\ref{sec:gaussian}), from nearby pixels (\S\ref{sec:sz3}), or with a neural network (\S\ref{sec:neural}).

Throughout this work, we report the cross entropy of the quantized value ${\rm bpp} = -\frac{1}{N}\sum_{i=1}^{N} \log_2 q_i$, where $N$ is the number of pixels and $q_i$ is the compression model's probability of the field value in pixel $i$, aside from when we use the SZ3 algorithm (\S~\ref{sec:sz3}), where we report the compressed file size that this algorithm outputs.  Although we do not build the coder ourselves, an arithmetic coder, for example, supplied with $q_i$ achieves this compression within a few bits  \citep{witten1987arithmetic}. For all three compression methods, the power spectrum is preserved at sub-percent level error (Appendix \ref{ap:pk}).

%Given the probability, numerous compression algorithms come close to this limit, such as arithmetic encoding \citep{witten1987arithmetic}.  

\subsection{Compression in the Gaussian limit}
\label{sec:gaussian}

In cosmology, statistical translational invariance means the Fourier transform diagonalizes the covariance, and the distribution of mode amplitudes at a given wavevector is nearly a zero-mean Gaussian: on linear scales from the initial conditions, and on nonlinear scales from the central limit theorem, since each mode sums over many roughly independent regions of the box. We therefore take the mode amplitudes to be independent Gaussians and ask what rate that implies.  This Gaussian limit also turns out to be where compressibility is perhaps best understood. 

Equation~(\ref{eqn:shannon}) is the rate for a fixed set of quantization bins, but for lossy compression we care about the minimum rate at a total tolerable error. We quantify this error by the mean square error distortion $D \equiv \langle (y(x)-x)^2 \rangle$, where $y(x)$ is the quantized value of $x$. Ref. \citep{shannon59} showed that for a Gaussian of variance $\sigma^2$ the minimum achievable rate is (c.f. Appendix~\ref{ap:Gaussianlimit})
\begin{equation}
{\rm bpp} = \left\{
\begin{array}{rl}
\frac{1}{2}\log_2\left(\frac{\sigma^2}{D}\right)
  & \mbox{if $\sigma^2 \geq D$}, \\
0 & \mbox{otherwise},
\end{array}\right.
\label{eqn:shannonlimit}
\end{equation}
the second condition expressing that once $D$ exceeds the variance no bits need be stored.

There is no reason to weight all Fourier modes equally, however, since each carries different power. Kolmogorov showed that the minimum achievable rate under this additional optimization at fixed distortion is:
\begin{equation}
{\rm \langle bpp \rangle} = N_m^{-1}\frac{1}{2}\sum_{\bfk}
  \log_2 \frac{P(k)}{D^*\langle P \rangle},
\label{eqn:Pkshannonlimit}
\end{equation}
where $N_m$ is the total number of Fourier modes, $P(k) \propto \langle|\delta_{\bfk}|^2\rangle$, and $\langle P \rangle \equiv N_m^{-1}\sum_{\bfk} P(k)$. Only modes with $P(k)/(D\langle P\rangle) > 1$ enter the sums, which for many of our cases is all of them; then $D^*$ is the RMS distortion $D$ and $N_m$ is half the number of real space pixels. Thus, it is advantageous to spend no bits on modes with $P(k)/(D\langle P\rangle) < 1$, and each such mode contributes to the variance in proportion to its power, so that $D^* = D - \langle P\rangle_c/\langle P\rangle$, where $\langle P\rangle_c$ sums over the excluded modes. 

We now apply this to our three types of simulations:
\begin{itemize}
\item For the 2D weak lensing images, the optimal Gaussian compression algorithm requires 4.1 and 0.97 bits at distortions of 0.01 and 0.1 against 6.6 and 3.3 if the same number of bits were allocated to each mode (so each mode had the same fractional distortion). %saving $2.7$ and $1.8$ bpp. 
The savings owes mostly to the gridding, which smooths on the grid scale and leaves substantially less power at the highest wavenumbers, making the high-wavenumber modes near or below the waterfilling level of eqn.~(\ref{eqn:Pkshannonlimit}).

%\item For the Bolshoi simulations 3D density grid,  we find $1.60$ and $0.37$ bits at the distortions of 0.01 and 0.1, which are well below the lensing values because high-wavenumber modes carry less power than the distortion budget and can be discarded. The density power spectrum falls roughly as $k^{-3}$ in 3D versus $k^{-2}$ in 2D, and there is additional suppression at high-$k$ due to cloud in cell interpolation onto the Eulerian grid.
\item For the Bolshoi simulations 3D density grid, we find $1.75$ and $0.43$ bits at the distortions of 0.01 and 0.1, which are well below the lensing values because high-wavenumber modes carry less power than the distortion budget and can be discarded. This is due to the cloud-in-cell interpolation used to place the Bolshoi particles onto the Eulerian grid, which suppresses high-$k$ power more than the gridding of the lensing maps does.

\item Next we consider the $N$-body case.  We find velocity components require $3.1$, $6.4$ and $9.7$ bits at $0.1$, $0.01$, $0.001$ distortion , and displacement requires $6.4$ bits at $0.001$. Storing velocity requires more bits than displacement because 90\% of the displacement variance lies below $k \approx 1\,h$/Mpc, whereas the velocity field reaches that fraction only near the Nyquist frequency. So water-filling can discard most of the displacement's modes but almost none of the velocity's.

%{\bf we should understand/comment on why velocity has more bits than displacement....it has to do with blueness of the power I think}

%This effect is stronger in 3D, where most modes lie near the Nyquist frequency and the density power spectrum falls roughly as $k^{-3}$ on the scales the grid resolves. 

%Therefore how much space gridding saves depends on the dimentionality and on simulations resolution limit based on the compression scheme.
 %This would improve over a fixed allocation of the same bits to per mode to meet these distortions on a per mode basis by XXXX. 

%The same scheme applies to the particle outputs of $N$-body simulations rather than the Eulerian gridded values just considered. Particles start on a regular grid and are displaced from it, so their Lagrangian coordinates are recoverable from the particle identifiers and the displacements and velocities form regular grids to which equation~(\ref{eqn:Pkshannonlimit}) applies unchanged.  

%The bpp for the velocity and displacement? are higher because their spectra are UV -- scaling as $k^{-1}$ relative to the density --, so essentially very few modes fall below the water level and almost every mode must be stored. 
\end{itemize}

We also compute, via the KL divergence, how far the distribution of each mode amplitude diverges from a Gaussian, and find this costs $<0.1$~bpp for the above fields. So the larger gains of non-gaussian models must therefore come from correlations between modes rather than from the modes individually.

%{\bf Maybe comment on how summary statistics like power spectrum here  are maintained (and say results apply to other algorithms discussed later).  This could also come later after you describe all algorithms.  For the gaussian compression, I think you expect the power spectrum error to decrease dramatically with decreasing k.  If you had the same bits per mode, the power spectrum of the error would scale as $k^{-3}$ and I think it's even better than this for the waterfilling}

\subsection{Interpolation based compression: SZ3}
\label{sec:sz3}

Interpolation-based compressors take the opposite approach to the Fourier scheme: rather than decorrelating the field globally, they predict each value from its already-decoded neighbors and store only the residual, which is small wherever the field is smooth on the pixel scale. SZ3 \citep{SZcompressor} adapts the predictor to the local structure of the data rather than assuming a fixed spectral form. It begins with low-order interpolation, a single point with a constant predictor, then two points at each end of a dimension with linear interpolation, and progressively increases the number of sample points and the spline order. It also predicts the value at one corner of a cube from the already-decoded values at the other corners, and for each region chooses whichever predictor produces the smaller residuals.  The residuals are quantized into bins of width twice the user-specified absolute error tolerance, encoded with Huffman coding, and passed through a lossless dictionary encoder that identifies repeated patterns. It has previously been used to explore extreme compression of cosmological datasets that maintain summary statistics by \citep{2020arXiv200400224J, 2021arXiv210400178J}; our exploration of percent-level pixel-scale distortions is much less aggressive.

Table~\ref{tab:gaussian} collects these rates alongside the Gaussian results of \S\ref{sec:gaussian}. Since SZ3's control parameter is a bound on the maximum error rather than the RMS error we use elsewhere, we vary the bound until the realized RMS error matches the target, so all rates in Table~\ref{tab:gaussian} are quoted at the same distortion. SZ3 requires more bits than the Gaussian coder on the lensing maps, but fewer on the Bolshoi density at $10^{-1}$ and $10^{-2}$ and on both Lagrangian fields, falling below even the Gaussian rate-distortion bound on the velocity. One of the reasons for this is smoothness on the real-space pixel scale, which lets a local predictor do well and explains the Lagrangian results. Also, a one-point distribution dominated by a single repeated value in the voids of the density is exploited by the Huffman and dictionary stages. 

%At the $10^{-4}$ precision of \S\ref{sec:numericalsims}, the Gaussian coder requires $9.7$ bits for the $x$ displacement against $9.5$ for SZ3. Holding the velocities to the same $10^{-4}$, the two require $13.1$ and $12.7$ bits. Ref.~\citep{guppy} report compression ratios consistent with SZ3 at the same accuracy.

%Ref. \citep{guppy} made the sensible recommendation for storing particle positions to a precision of one force softening length, which for our grid and softening length of $1/50^{\rm th}$ the mean inter-particle spacing corresponds to a distortion of $2\times10^{-4}$ in units of the field's standard deviation. At that precision SZ3 stores the displacement in $7.3$ bits per component against $8.5$ for the Gaussian coder, and Ref.~\citep{guppy} report compression ratios consistent with SZ3 at the same accuracy.

\begin{table*}
\centering
\caption{Bpp at the specified RMS error. `Bound' is the Shannon limit for a Gaussian field of the same power spectrum, $N_m^{-1}\frac{1}{2}\sum_{\bm k}\log_2(P_k/\theta)$ over modes above the water level $\theta$. `Coder' is achievable: each retained mode uniformly quantized and priced by its Gaussian bin probability, with sub-threshold modes zeroed, which the decoder identifies from $P(k)$ without side information. The two differ by the scalar space-filling loss, $\frac{1}{2}\log_2(2\pi e/12) \approx 0.25$~bpp when every mode is retained and less when modes are discarded \citep{GishPierce1968}. SZ3 rates are also achievable. Its control parameter is an absolute error bound, which we adjust for each field until the realized RMS error matches the target. A non-Gaussian field can be compressed below the Gaussian bound \citep[Theorem 4.63]{doi:https://doi.org/10.1002/0471219282.eot142}, which SZ3 achieves on the velocity at every distortion and on the Bolshoi density at $10^{-1}$ and $10^{-2}$. Displacement rates are given at $10^{-3}$ only, the precision discussed in \S\ref{sec:numericalsims}. }
\label{tab:gaussian}
\small
\setlength{\tabcolsep}{4pt}
\begin{tabular}{lcc ccc ccc ccc}
\hline
 & & &  \multicolumn{3}{c}{$10^{-1}$} & \multicolumn{3}{c}{$10^{-2}$}
     & \multicolumn{3}{c}{$10^{-3}$} \\

simulation & description & $N_{\rm grid}$ 
  & \shortstack{Gauss.\\bound} & \shortstack{Gauss.\\coder} & SZ3
  & \shortstack{Gauss.\\bound} & \shortstack{Gauss.\\coder} & SZ3
  & \shortstack{Gauss.\\bound} & \shortstack{Gauss.\\coder} & SZ3 \\
\hline
\multicolumn{12}{l}{\textit{Eulerian grids}} \\
ColombiaLens & lensing $\kappa$ & $1024^2$
  &  0.775 & 0.969 & 1.40 & 3.87 & 4.14 & 4.51
  & 7.19 & 7.44 & 7.98 \\
% & & & log & 0.762 & 0.967 & 1.57 & 3.87 & 4.17 & 4.77
%  & 7.19 & 7.47 & 8.17 \\
\addlinespace
Bolshoi Grid & CIC+GS & $512^3$
  &  0.367 & 0.429 & 0.175 & 1.60 & 1.75 & 1.27
  & 3.82 & 4.04 & 4.16 \\
% & & & log & 0.149 & 0.180 & 0.604 & 1.32 & 1.52 & 3.44
%  & 4.27 & 4.55 & 6.90 \\
\hline
%\multicolumn{13}{l}{\textit{Lagrangian}} \\
%GADGET & DM $x$-displ.$^\dagger$ & $512^3$
%  & linear & --- & --- & --- & 5.0 & --- & 3.0 & 8.3 & --- & 6.1 \\
%GADGET & DM velocity & $512^3$
%  & linear & 2.4 & --- & 1.9 & 5.7 & --- & 5.1 & 9.1 & --- & 8.4 \\
\multicolumn{12}{l}{\textit{Lagrangian}} \\
CAMELS & DM $x$-displ. & $256^3$
  &   &  &  &  &  & 
  & 6.11 & 6.37 & 6.13 \\
CAMELS & DM $x$-velocity & $256^3$
  &  2.83 & 3.10 & 2.60 & 6.15 & 6.41 & 5.88
  & 9.48 & 9.73 & 9.23 \\

\hline
\end{tabular}
\end{table*}

\subsection{Neural compression}
\label{sec:neural}

We now use a neural network to learn the full joint probability of each value from the field being compressed, so it is not restricted to the Gaussian ansatz nor to only nearby smooth correlations, as in the previous algorithms. 
%A non-Gaussian field  can always be compressed more than a Gaussian field with the same power spectrum \citep{cover2006elements}.
%% add other method issues

Previous work in neural image compression is rooted in learned transform coding \citep{2016arXiv161101704B, 2022arXiv220206533Y}, in which an encoder, a quantizer and an entropy model are jointly trained to trade bit rate against distortion. We implemented a similar approach and found the tradeoff difficult to control, with us typically unable to create networks that would achieve distortions of $\lesssim 10^{-2}$. Since our compressor is specific to scientific data where such small distortions are generally necessary, we instead fix the quantizer and learn the probability model. 

Quantizing uniformly in the field values with step $\delta = \sqrt{12}D^{1/2}$ yields the target root-mean-square distortion $D^{1/2}$ by construction, for any model and any state of training. The rate is then set entirely by how well we can learn the conditional probability $p(i_n|i_{<n})$ that the $n^{\rm th}$ pixel falls in bin $i_n$ given the previously coded bins $i_{<n} \equiv \{i_0,\ldots,i_{n-1}\}$:
\begin{equation}
{\rm optimal~bpp} = -\sum_{i_n} p(i_n|i_{<n}) \log_2 p(i_n|i_{<n}).
\label{eqn:autoregressive}
\end{equation}
The model factorizes the image at two scales. A transformer \citep{2017arXiv170603762V} over $32\times32$-pixel patches in raster order supplies long-range context. Each patch is summarized by a context vector conditioned on all preceding patches. Within each patch, masked convolutions condition each pixel on the previously decoded pixels of its own patch and on the patch context vector. For each pixel the model outputs a probability density for its value, a mixture of eight Gaussians whose parameters are set by the pixels already decoded \citep{Salimans2017}. Integrating this density over the pixel's quantization bin gives the model's estimate of $p(i_n|i_{<n})$. An arithmetic coder spends $-\log_2$ of this probability in bits on the pixel, so the better the model predicts each pixel, the fewer bits it needs. A decoder working through the field in order has only the quantized values of earlier pixels, so we compute the mixture from those rather than the original values, which makes the reported rate achievable in practice. Training never uses the quantizer: the model maximizes the density it assigns to the unquantized pixels, which does not depend on $\delta$, so one trained model can be evaluated at any $\delta$ and prices the entire rate-distortion curve. Because the distortion is fixed by the quantizer, an imperfectly trained model costs bits but never accuracy.

Extending this to 3D, requires only that the patches become three-dimensional. We dice a $128^3$ sub-cube into small cubes of $8\times8\times8 = 512$ voxels, each of which becomes a single token, and masked three-dimensional convolutions within a patch condition each voxel on the voxels of that patch already decoded. The same architecture applies to the Lagrangian fields, where the grid index is the particle's initial position rather than its position at $z=0$.

Table~\ref{tab:compression} highlights the results of compressing our simulation fields with this model. Uniform scalar quantization carries an irreducible penalty of $\frac{1}{2}\log_2(2\pi e/12) \approx 0.25$ bpp relative to the rate-distortion bound at the same mean squared error \citep{GishPierce1968, gersho_gray_1992}, paid before the probability model enters. The Gaussian coder of \S\ref{sec:gaussian}, which uniformly quantizes each retained Fourier mode and prices it by its Gaussian bin probability, pays this same penalty, so it rather than the bound is the fair comparison. Against this, the conditional model has access to the non-Gaussian structure of the field, in particular the phase correlations that the mode-by-mode Gaussian treatment discards. The conditional gain wins on every field we considered, and by the largest margin where the field is most non-Gaussian: the transformer requires 3.58 bpp on the lensing maps against 4.14 for the Gaussian coder, and 1.28 against 1.75 on the three-dimensional density field. The converse also holds. On log of the Bolshoi density, which is close to Gaussian, the lensing-trained model does not beat the Gaussian coder since reverse water-filling is already near optimal there and leaves little non-Gaussian structure for a learned prior to exploit. It also falls below the Gaussian rate-distortion bound on every field, which a non-Gaussian field is permitted to do \citep[Theorem 4.63]{doi:https://doi.org/10.1002/0471219282.eot142}. On the full density box SZ3 reaches 1.27 bpp, level with the transformer, but restricted to the same $128^3$ sub-cubes the transformer sees it needs 1.44.

We also find that the model transfers between fields. Applied to 2D slices of the 3D $N$-body density with no retraining at all, a model trained only on lensing maps reaches 2.65 bpp against 2.81 for the Gaussian coder, and fine-tuning on a small subset of the box brings it to 2.16. That a model trained on a much different dataset (the lensing maps only have percent-level variations, whereas for the 3D density the variations are factors of a hundred) already performs sensibly suggests it has learned generic properties of cosmological structure rather than features of its training set. In practice this makes compressing a new field simple: rather than training from scratch, one can fine-tune a pretrained model on a small portion of the new field, even when it looks quite different from the training data. We give details in Appendix~\ref{ap:transfer}.

\begin{table}
\centering
\caption{Bpp for each method, at $10^{-2}$ RMS error unless marked. All three columns are achievable rates. Rows marked $^*$ are at $10^{-3}$ distortion for particle displacements.}
\label{tab:compression}
\begin{tabular}{llccc}
\hline
field & $N_{\rm grid}$ & \shortstack{Gauss.\\coder} & SZ3 & Transformer \\
\hline
lensing $\kappa$              & $1024^2$ & 4.14 & 4.51 & 3.58 \\
Bolshoi density               & $512^3$  & 1.75 & 1.27 & 1.28 \\
CAMELS $x$-velocity           & $256^3$  & 6.41 & 5.88 & 5.29 \\
CAMELS $x$-displ$^*$   & $256^3$  & 6.37 & 6.13 & 5.77 \\
\hline
\end{tabular}
\end{table}

We apply the same fixed-quantization approach to the Lagrangian output of a CAMELS $N$-body run. These are the fields where SZ3 is strongest, beating the Gaussian coder on both and even the Gaussian rate-distortion bound on the velocity. For these fields we use the lensing-trained model fine-tuned on two-dimensional slices of the box (Appendix~\ref{ap:transfer}), rather than a three-dimensional model as for the density. The transformer still beats both: 5.77 bpp for the $x$ displacement at $10^{-3}$ distortion against 6.13 for SZ3 and 6.37 for the Gaussian coder, and 5.29 for a velocity component at $10^{-2}$ against 5.88 and 6.41. At $10^{-4}$ distortion it needs 9.19 bpp for the $x$ displacement against 9.47 for SZ3 and 9.69 for the Gaussian coder, and 11.48 for the $x$ velocity against 12.70 and 13.05. Its saving over the Gaussian coder on the displacement changes little between $10^{-3}$ and $10^{-4}$, 0.60 and 0.50 bpp, as expected in the high-resolution limit.

Storage has shaped the output format of only a few large simulation suites \citep{guppy}, the largest being AbacusSummit, with $6\times10^{13}$ particles. Its full particle time slices store positions and velocities at 12 bits per component relative to a per-cell scaling, with $0.3\,h^{-1}$\,kpc position granularity, then compress these losslessly to an effective 11.3 bits per component \citep{2021MNRAS.508.4017M}. At the $10^{-3}$ and $10^{-4}$ of \S\ref{sec:numericalsims}, the transformer needs 5.8 and 9.2 bits per displacement component, and 5.3 per velocity component at $10^{-2}$.  AbacusSummit also stores particle IDs in separate files, whereas on our Lagrangian grid each particle's identity is simply its place in the array. For analyses that do not need particle identities, a cell-based format like AbacusSummit's could equally be coded with a learned model.

\section{Conclusions}
\label{sec:conclusion}

Cosmological simulation outputs are stored at far higher precision than any analysis requires. Percent-level accuracy at the pixel scale, already conservative given the numerical and astrophysical uncertainties involved, costs a few bpp rather than the 32 typically stored. In this work, we measured what those few bits are by benchmarking achievable compression across weak lensing convergence maps, gridded $N$-body densities, and the particle displacements and velocities of an $N$-body run. We compared a Fourier-space scheme optimal in the Gaussian limit, the state of the art scientific compression tool SZ3, and our autoregressive transformer that fixes the quantizer and learns the probability model.  In general, we found 10-30\% differences between all the algorithms, which we also think suggests that we are pushing up against the maximal compression of these data sets.

The Fourier scheme, which allocates bits among modes by reverse water-filling under the assumption that the mode amplitudes are independent Gaussians, reaches percent-level accuracy at 1.8--6.4 bpp, depending on the numerical dataset. These would represent a $5$--$18\times$ compression of the standard 32-bit float. SZ3 beats the Gaussian coder on the 3D density with 1.3 bpp compared to $1.8$ at $10^{-2}$. On the velocity the improvement is again about half a bit, $5.9$ compared to $6.4$, and a quarter of a bit on the displacement, but not on the 2D lensing maps that have less spatial coherence and are more Gaussian, where the Gaussian coder prevails over SZ3 at $4.1$ relative to $4.5$. In comparison, fixing the quantizer and learning only the probability of each bin with an autoregressive transformer gives the lowest bit rate on every field in Table~\ref{tab:compression}, with the density compared within the same $128^3$ sub-cubes. It is 9 to 27\% below the Gaussian coder and 6 to 21\% below SZ3.

%The margin is largest where the field is most non-Gaussian.

%, and the penalty for that assumption is of order a tenth of a bit. 

%, so an interpolating predictor helps only where the field is smooth in real space or its values pile up at a single point. 
%How much the non-Gaussian structure is worth therefore depends strongly on the variable being coded, and both non-Gaussian methods we tried can compress below the Gaussian rate-distortion bound, which a non-Gaussian field is permitted to do. 

%We report cross entropies throughout rather than building the coder itself. An arithmetic coder attains these to within a negligible overhead, so the rates are real, but the autoregressive decoder is sequential and would run far slower than SZ3. Our uniform scalar quantizer also gives up roughly $0.25$ bpp to the rate-distortion bound, so a vector quantizer would improve every rate reported here by a comparable amount. Neither limits what is achievable in principle.

Our same approach should apply to observational data, where the storage problem is at least just as acute. The Vera C. Rubin Observatory will generate about 20 terabytes of raw images per night and roughly 60 petabytes over the ten-year Legacy Survey of Space and Time, with several hundred petabytes once processed \citep{2019ApJ...873..111I}, and Euclid faces a comparable challenge. At these volumes, storing, transferring and reanalyzing the data are all limited by the same bottleneck. On Rubin Data Preview 1 \citep{10.71929/rubin/2570308} coadds quantized at a quarter of the background noise, our fine-tuned transformer needs 4.0 bpp, 27\% below the Gaussian coder and 7\% below SZ3 as shown in \ref{app:dp1}. However, because observations generally store more-compressible integer counts, the advantages over a lossless encoder are smaller compared to cosmological simulations.

Finally, lossy compression is intriguingly connected to the fundamental limits of cosmological information extraction.  The number of bits required to store a field to a set precision relates to how information is lost in inevitably-imperfect measurements of large-scale structure. Moreover, optimal compression requires knowledge of the full probability distribution of a simulation's fields, just as optimal parameter inference requires understanding how this probability distribution depends on cosmological parameters. Thus, it might be possible to connect the limits of lossy compression to a noisy field's cosmological parameter constraining power.

\paragraph{Acknowledgments}
We would like to thank Lehman Garrison for helpful conversations, and Zoltan Haiman for helping us download the lensing simulations.  We acknowledge support from NASA grant 80NSSC24K1220.

\bibliography{References}

\appendix{}

\section{Transfer between fields}
\label{ap:transfer}

The autoregressive model is trained on the field it compresses, so it is worth asking how much of what it learns is specific to that field. We take the model trained on the ColumbiaLens convergence maps and apply it, with no retraining, to two-dimensional slices of the Bolshoi density and the CAMELS displacement. Each slice is standardized to zero mean and unit variance and tiled into $32\times32$ patches, matching the input format the model was trained on.

On the Bolshoi density the lensing-trained model reaches 2.65 bpp at $10^{-2}$ distortion against 2.81 for the Gaussian coder, so a model that has never seen a three-dimensional density field prices it below a scheme built from that field's own power spectrum. On the displacement at $10^{-3}$ it needs 6.82 against 6.37. Fine-tuning on slices from one side of the box and evaluating on slices from the far side gives 2.16 and 5.77 bpp, both below the Gaussian bound (2.62 and 6.11). The fine-tuned displacement model is the one reported in Table~\ref{tab:compression}. The density rates here are for two-dimensional slices for both methods, so they sit well above the $128^3$ sub-cube rates of Table~\ref{tab:compression}.

Zero-shot transfer is asymmetric. A model trained on the Bolshoi density needs 14.3 bpp on the lensing maps against 4.14 for the Gaussian coder. Standardized, the ColumbiaLens maps run from $-2.1$ to $+41$, while the Bolshoi density spans $-0.18$ to $+353$, bounded below by an empty cell. Over half of all lensing pixels therefore fall outside the range a Bolshoi-trained prior ever represented, while the Bolshoi values lie inside the range lensing covers. Fine-tuning on 16 lensing maps nonetheless recovers 3.81 bpp, against 3.58 for a model trained on lensing throughout.

The zero-shot rate therefore measures how far a prior sits from useful on a new field, and which direction succeeds is consistent with the one-point distributions. In both directions fine-tuning on a small fraction of the target brings the rate well below its zero-shot value, so adapting a pretrained model to a new simulation is cheap compared with training one from scratch, and what the model learns is not specific to the field it was trained on.

\section{Transfer learning for observational data: Rubin DP1}
\label{app:dp1}

For this analysis, we used Rubin DP1 data. The simulation rows of Table~\ref{tab:gaussian} fix the acceptable distortion of the field standard deviation. That convention does not carry over to survey images. Here we take coadd images from Rubin which are background-subtracted, so most pixels sit near zero and a fractional tolerance is meaningless there. The field standard deviation is set by whichever sources happen to fall in the cutout rather than by anything intrinsic.

The natural reference is instead the noise already present in the pixel. Ref. \citep{2010PASP..122.1065P} quantize floating-point images onto levels spaced at $\delta = \sigma/q$, where $\sigma$ is the measured background noise and $q$ is how many quantization levels span one unit of noise i.e. q = 4 means the grid is four times finer than the noise.

\begin{equation}
  \sigma_q^2 = \sigma_0^2 + \delta^2/12 ,
\end{equation}
so the quantization error variance is $\delta^2/12$. Substituting $\Delta=\sigma/q$,
\begin{equation}
  D = \frac{\sigma^2}{12\,q^2},
  \qquad
  \sqrt{D} = \frac{\sigma}{q\sqrt{12}} .
  \label{eq:dp1target}
\end{equation}

The quantizer step $\delta = \sqrt{12 D}$. We measure $\sigma$ from the pixels themselves using \citep{2010PASP..122.1065P}, a third-order median absolute difference

\begin{equation}
  \sigma = 0.6052 \times \mathrm{median}\left(\left|\,
  -x_{i-2} + 2x_i - x_{i+2} \,\right|\right),
  \label{eq:mad}
\end{equation}
evaluated along image rows. This choice keeps the target free of any assumption about the survey's own noise model. Because this samples every other pixel, a noise correlation must extend over at least two pixels to bias it. We apply no source masking.
\begin{table}[h]
\centering
\caption{bpp on DP1 coadd $512^2$ stamps. The $q$ rows use Eq.~\ref{eq:dp1target} with $\sigma$ from Eq.~\ref{eq:mad}, evaluated per stamp. All rates are achievable.}
\label{tab:dp1}
\begin{tabular}{lcccccc}
\hline
target &  Gauss.\ coder & SZ3 &
NN zero-shot & NN fine tuned \\
\hline
$q=1$           &   2.85 & 2.33 & 5.65  & \textbf{2.10} \\
$q=4$         &   5.530 & 4.322 & 8.657  & \textbf{4.03} \\
$q=16$              &  7.59 & 6.37 & 10.87 & \textbf{6.03} \\
\hline
\end{tabular}
\end{table}

Ref. \citep{2010PASP..122.1065P}  quantized simulated and real CCD images and measured the degradation in \textsc{SExtractor} magnitudes and positions directly: at $q=4$ the background noise rises by $0.26\%$ and the magnitude and position uncertainties of faint stars by $0.31\%$ and $0.18\%$; at $q=1$ the corresponding figures are $4.1\%$, $5.6\%$ and $4.1\%$. They recommend $q$ between $1$ and $4$. We therefore report $q=1$, $4$ and $16$, bracketing that range with $q=4$ as the default.

We find similar performances as simulations: the fine-tuned transformer gives the lowest rate at every target, better than SZ3 and Gaussian coder. The margin over the Gaussian coder is $21$ -$27\%$, comparable to the largest gaps in Table~\ref{tab:compression}. The rates scale as expected with the quantizer: each factor of four in $q$ costs very nearly two bpp ($1.93$ and $2.00$ for the fine-tuned model), confirming that the three operating points differ only in the grid and not in what the model has learned.

Zero-shot transfer using the model trained on the lensing maps is poor, at $5.65$--$10.87$ bpp, and fine-tuning brings it to $2.10$--$6.03$ bpp. Sky-subtracted coadds are two-sided and contain bright compact sources, so a lensing-trained prior is asked to price values well outside the support it was trained on. As elsewhere, the cost is in bits and not accuracy, since the distortion is fixed by the grid.

At the recommended $q=4$ the fine-tuned rate is $4.03$ bpp against the $32$ stored. For comparison, Ref.\citep{2010PASP..122.1065P} predict $\log_2 q + 1.792 + K$ bpp for quantized noise alone, which with their $K\simeq1.2$ gives $4.99$ at $q=4$. The learned model therefore prices the entire image, sources included, below what a production lossless coder spends on the noise by itself. 

The advantage over lossless coding is still less dramatic than for cosmological simulations as raw 16-bit CCD frames are integer counts clustered about the sky background, so the best lossless coders already reach as few as 5 bpp \citep{2025arXiv250608306T}, about one bit above our $q=4$ bpp. Simulations stored as 32-bit floats have no such head start, since all but the first few significant bits of each value are effectively noise and cannot be removed losslessly.

\section{Power Spectrum preservation}
\label{ap:pk}

\begin{figure}
\centering
\includegraphics[width=.5\columnwidth]{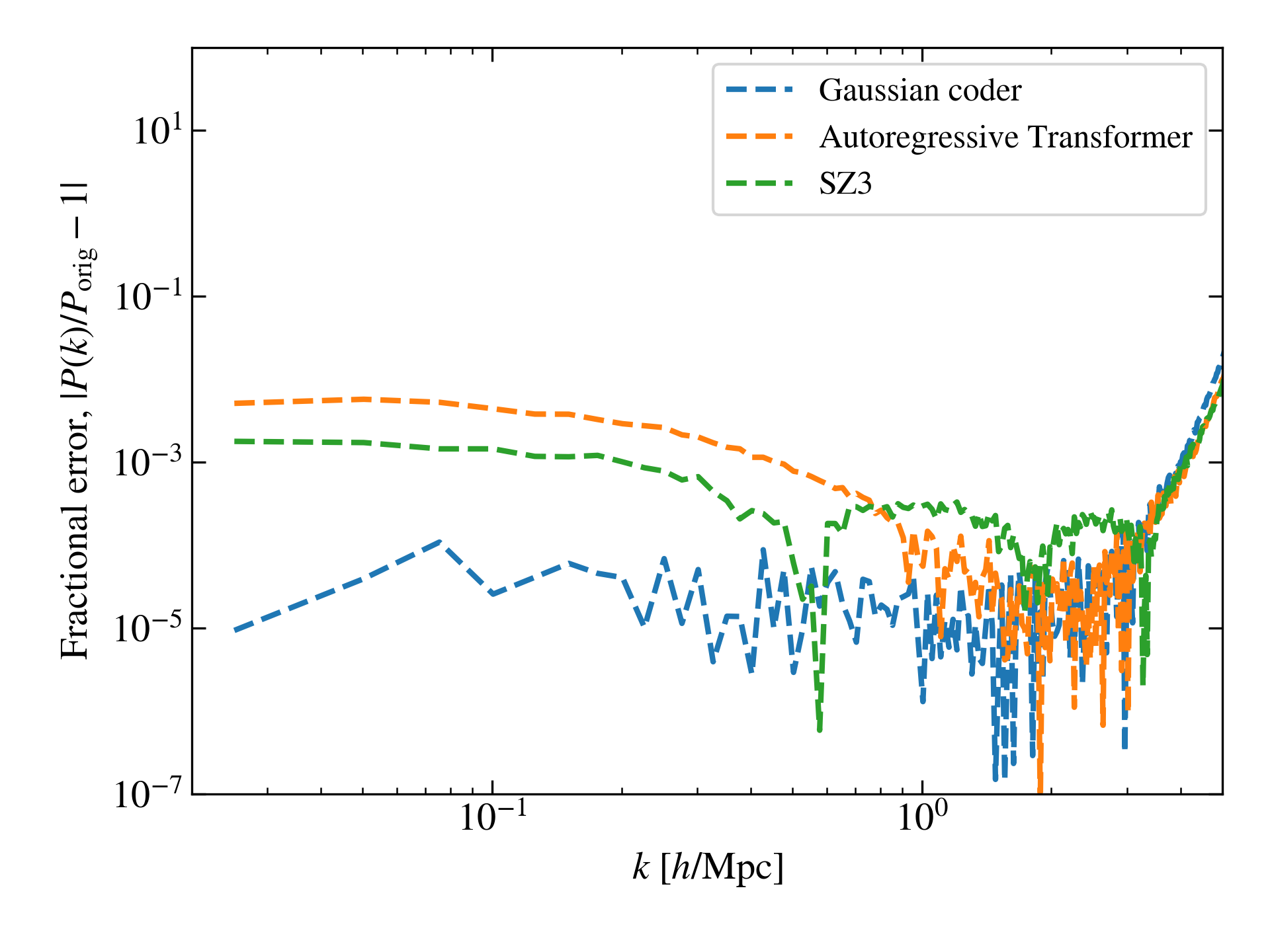}
\caption{Fractional error in the power spectrum of the 3D Bolshoi density field after compression at $10^{-2}$ RMS distortion, for the Gaussian coder, SZ3 and the autoregressive transformer. Wavenumbers are in units of $h$ Mpc$^{-1}$, where $h$ is the Hubble parameter in units of $100~\mathrm{km~s^{-1}~Mpc^{-1}}$. All three methods preserve $P(k)$ to better than a percent across every scale the grid resolves.}
\label{fig:pk}
\end{figure}

Percent-level pixel-scale error is only useful if it does not bias the statistics the data are used for. We compress the 3D Bolshoi density field at $10^{-2}$ RMS distortion with each method and compare the power spectrum, $P(k)$, of the reconstruction against the original in Figure~\ref{fig:pk}.  The power spectrum is the favored summary statistic in cosmology, and needs to be maintained to at least 1\% precision at $k \lesssim 1~h/$Mpc to not bias the percent level determinations of cosmological parameters coming from surveys \citep{2010ApJ...715..104H}.

All three methods preserve $P(k)$ to better than 1\% over the full range of scales the grid resolves. The fractional error grows above $k\approx1\,h$/Mpc, well beyond the $k\lesssim1\,h$/Mpc range where percent-level accuracy is required. The grid resolves to $k_{\rm Nyq}=6.4\,h$/Mpc, and the CIC window suppresses power by a factor of several approaching that scale, so the fractional error rises because the denominator is falling.

The Gaussian coder is the most faithful, flat to within $10^{-4}$ across the resolved range. It works mode by mode and spends more bits on the largest modes, which carry the most power, so the fractional error is small everywhere. The two real-space methods carry small biases on large scales, the transformer by about half a percent and SZ3 by about 0.2\%. The transformer and SZ3 fall below $10^{-3}$ by $k\approx0.5\,h$/Mpc.

\section{The rate-distortion function of a Gaussian}
\label{ap:Gaussianlimit}

Here we sketch the derivation of equation~(\ref{eqn:shannonlimit}), following \citet{cover2006elements}. The rate-distortion function is the smallest mutual information between the source $X$ and its reconstruction $\hat{X}$ consistent with the distortion constraint,
\begin{equation}
R(D) = \min_{\langle (\hat{X}-X)^2 \rangle \le D} I(X;\hat{X}),
\end{equation}
and Shannon's theorem states that this rate is achievable, and no lower rate is, in the limit of long block lengths.

Taking $X \sim \mathcal{N}(0,\sigma^2)$ and letting $\hat{X}$ be any reconstruction satisfying the constraint and writing the mutual information in terms of differential entropies and using $h(X|\hat{X}) = h(X - \hat{X}|\hat{X})$,
\begin{eqnarray}
I(X;\hat{X}) &=& h(X) - h(X|\hat{X}) \nonumber \\
 &=& \tfrac{1}{2}\log_2(2\pi e \sigma^2) - h(X - \hat{X}\,|\,\hat{X})
   \nonumber \\
 &\ge& \tfrac{1}{2}\log_2(2\pi e \sigma^2) - h(X - \hat{X}) \nonumber \\
 &\ge& \tfrac{1}{2}\log_2(2\pi e \sigma^2)
   - \tfrac{1}{2}\log_2\!\left(2\pi e \left\langle (X-\hat{X})^2
     \right\rangle\right) \nonumber \\
 &\ge& \tfrac{1}{2}\log_2(2\pi e \sigma^2)
   - \tfrac{1}{2}\log_2(2\pi e D)
 \;=\; \tfrac{1}{2}\log_2\frac{\sigma^2}{D}.
\label{eqn:rdlower}
\end{eqnarray}
The three inequalities use, in order, that conditioning cannot increase entropy; that among all distributions of a given variance the Gaussian has the largest differential entropy; and the distortion constraint $\langle (X-\hat{X})^2 \rangle \le D$.

The bound is attained. Consider the Gaussian test channel
\begin{equation}
\hat{X} = \sqrt{1 - D/\sigma^2}\;X + Z, \qquad
Z \sim \mathcal{N}(0, D)\ \ \mathrm{independent\ of}\ X,
\end{equation}
for $D < \sigma^2$. A short calculation gives $\langle (X - \hat{X})^2 \rangle = D$ and
$I(X;\hat{X}) = \tfrac{1}{2}\log_2(\sigma^2/D)$, so equation~(\ref{eqn:rdlower}) holds with equality. If instead $D \ge \sigma^2$, setting $\hat{X}=0$ meets the constraint at zero rate, which gives the second branch of equation~(\ref{eqn:shannonlimit}).

Two features of the optimal scheme are worth noting, since they explain why practical coders do not reach this rate. The reconstruction is shrunk toward the mean by $\sqrt{1-D/\sigma^2}$ rather than being an unbiased estimate of $X$, and independent noise is then added; it is not a deterministic quantizer. Achieving $R(D)$ also requires coding long blocks jointly. A scalar quantizer followed by entropy coding, which is what we and the compression packages we compare against actually implement, pays a further $\tfrac{1}{2}\log_2(2\pi e/12) \approx 0.25$ bits per sample in the high-resolution limit \citep{GishPierce1968}, the difference between tiling space with hypercubes and with the optimal cell shape.

\end{document}